# Performance of the solid deuterium ultra-cold neutron source at the pulsed reactor TRIGA Mainz

J. Karch[1], Yu. Sobolev[2,a], M. Beck[1], K. Eberhardt[2], G. Hampel[2], W. Heil[1,b], R. Kieser[1], T. Reich[2], N. Trautmann[2], and M. Ziegner[3]

[1] Institute of Physics, Johannes Gutenberg-Universität Mainz, Staudingerweg 7, D-55128 Mainz, Germany

[2] Institute of Nuclear Chemistry, Johannes Gutenberg-Universität Mainz, Fritz-Strassmann-Weg 2, D-55128 Mainz, Germany

[3] Austrian Institute of Technology GmbH, Health & Environment Department, Muthgasse 11, 1190 Wien, Austria

**Abstract.** The performance of the solid deuterium ultra-cold neutron source at the pulsed reactor TRIGA Mainz with a maximum peak energy of 10 MJ is described. The solid deuterium converter with a volume of V=160 cm$^3$ (8 mol), which is exposed to a thermal neutron fluence of $4.5 \times 10^{13}$ n/cm$^2$, delivers up to 550 000 UCN per pulse outside of the biological shield at the experimental area. UCN densities of ~ 10/cm$^3$ are obtained in stainless steel bottles of V ~ 10 L resulting in a storage efficiency of ~20%. The measured UCN yields compare well with the predictions from a Monte Carlo simulation developed to model the source and to optimize its performance for the upcoming upgrade of the TRIGA Mainz into a user facility for UCN physics.

## 1. Introduction

Ultra-cold neutrons (UCN) offer unique opportunities for investigating the properties of the free neutron with exceptionally high precision. UCN are defined as neutrons that can be trapped in material bottles or confined by magnetic potential walls. The high-energy limit of UCN is set by the neutron optical potential (Fermi potential) of the material selected, e.g., 342 neV in case of $^{58}$Ni, or by the magnetic barrier acting upon the neutron's magnetic moment ($|\mu_n| = 60$ neV/Tesla) in an inhomogeneous magnetic field of a magnetic bottle.

Long observation times are the key feature of experiments with UCN enabling high precision measurements which allow the search for new physics becoming manifest itself as small deviations from expectations predicted by the Standard Model. For instance, the ambitious aim for the next generation of experiments [1-6] is the search for an electric dipole moment (EDM) of the neutron ($d_n$) with a sensitivity in the range of $\delta d_\text{n}$ : [$10^{-27}$ ecm - $10^{-28}$ ecm].

---

[a] on leave from: Petersburg Nuclear Physics Institute, Gatchina, Russia.

[b] corresponding author: wheil@uni-mainz.de

The physics interest behind the *EDM experiment* is related to a possible observation of the breaking of a fundamental symmetry in particle physics, the CP symmetry, which may explain the asymmetry of matter over antimatter in the universe. This symmetry breaking is not yet understood and belongs to the most exciting topics in particle physics, both experimentally and theoretically.

Trapped neutrons also allow to test the structure of weak interaction: Neutron ß-decay offers a number of independent observables, considerably larger than the small number of parameters describing this decay in the Standard Model. While the neutron lifetime gives the overall strength of the weak interaction, neutron decay correlation coefficients depend on the ratios of the coupling constants involved, and hence determine its internal structure [7]. Today, all semi-leptonic charged current weak interaction cross sections needed in cosmology, astrophysics, and particle physics must be calculated from neutron decay parameters [8]. With the ongoing refinement of models of cosmology, astrophysics, and of particle physics, the growing requirements on the precision of these neutron decay data must be satisfied by new experiments.

The statistical limitation in all these storage measurements is the maximal obtainable number of UCN. The conventional production mechanisms are reactor based and rely on extracting the lowest-energy part of the neutron spectrum, which in most cases may be assumed to be approximately Maxwellian with a characteristic temperature $T_m$ of the moderator. The use of a moderator cooled to liquid deuterium temperature (25 K, cold source) increases the maximum density available from thermal reactors by a factor of $(300/25)^{3/2} \approx 40$ [9]. The UCN flux can be increased further by either vertical extraction from the cold moderator and/or by mechanical deceleration which uses a moving scatterer carrying away the neutron momentum. These UCN production methods are based on the actions of conservative fields, under which the phase space volume is invariant. This limits the UCN intensity produced by sources of this kind.

The superthermal UCN mechanism may produce UCN densities orders of magnitude greater than existing thermal UCN sources. It was proposed first in 1975 by Golub and Pendlebury [10]. In a superthermal source, a neutron dissipates its energy by exciting the collective modes of condensed matter under investigation that carry away the entropy of the neutron flux. If the downscattering rate is much larger than the upscattering rate, then a large quantity of UCN can be accumulated. In principle, the density is limited by the nuclear absorption time or the neutron lifetime. The idea was first pursued with superfluid $^4$He [11] and only in the past view years the feasibility of such an UCN source was successfully demonstrated [12,13].

Materials other than $^4$He can very efficiently downscatter cold neutrons into UCN but are limited by their nuclear absorption probability. Solid deuterium (sD$_2$) is one such material. To compensate for the short production time, first pointed out by Golub and Boning [14], a small volume of sD$_2$ is coupled to a large inert storage volume. In this way, production and storage of UCN is separated. Pokotilovski pointed out the advantages of UCN production with sD$_2$ at pulsed neutron sources like TRIGA reactors [15]. The use of spallation as a pulsed source for neutrons was suggested by the Gatchina group [16] and is verified among others at PSI [17]. Superthermal UCN sources are now in operation or being under construction at different places worldwide, such as PSI (Villigen) [17], Los Alamos [18], ILL (Grenoble) [12], RCNP, Osaka University [13], PULSTAR reactor (NC State University) [19], FRM-II (Garching) [20] , and TRIUMF [6] .

Low-power reactors, such as the TRIGA Mainz, are competitive due to the possibility to pulse the reactor every five minutes and to produce a high density of UCN in the pulse that ideally meets the requirements of storage experiments, where the trap must be filled with a similar frequency. Here, we report on results that have been obtained with a pulsed UCN source at the radial beam port D of the TRIGA reactor at Mainz University. The source moderates and converts thermalized neutrons to UCN in a windowless sD$_2$ volume inside of a UCN guide system. The neutrons are transported to an experimental area outside of the biological shield. The design of the source as well as the UCN production under operational conditions are described and  numbers for UCN densities obtained in material bottles of V~ 10 L , typically, are presented. Finally, a detailed analysis of the measured time-of-flight spectra in connection with Monte Carlo (MC) simulations are given, indicating the route for a further upgrade of the source to a targeted strength of ~50 UCN/cm$^3$.

## 2. The TRIGA Mainz UCN source

The inherent safety of TRIGA reactors is due to a physical property of the fuel elements yielding a large prompt negative temperature coefficient [21-23]. For pulse mode operation, the TRIGA Mainz reactor is brought to criticality at a low steady-state power (50 W) and then a control rod is shot out of the reactor core by compressed air. Due to this sudden insertion of excess reactivity, the power rises sharply with a reactor period of only a few milliseconds. At the TRIGA Mainz reactor, a maximum peak power of 250 $MW_{th}$ in the pulse mode is obtained with a pulse width at half maximum (FWHM) of about 30 ms. The maximum pulse rate is 12 per hour, which matches the duty factor of typical UCN storage experiments. Based on the experiences with a prototype UCN source installed at the tangential beamtube C at the TRIGA Mainz [24], a new superthermal UCN source at the radial beamtube D was designed and tested. Details of the source are shown schematically in Figure 1. Major developments of the new source are mainly directed to the optimization of the UCN converter volume, the reduction of the consumption of liquid helium (8 L/h), necessary for cooling the converter to its work temperature of $\approx$ 6 K, and the implementation of an efficient premoderator for neutrons. Furthermore, a 10 L buffer volume was inserted acting both as *l*He reservoir and as phase separator (gas/liquid) in between the *l*He inlet line and the cooling pipes to the converter nose. The technical aspects of the source are discussed in detail in ref. [25]. Thus, only a short description is given: The cryostat consists of a vessel outside the reactor shielding and the in-pile part. The in-pile part of the cryostat houses the UCN guide, an electro-polished uncoated stainless steel tube with an internal diameter $\varnothing_{in}$ = 66 mm and a total length of about 4.4 m[1]. At the position of the vertical cryostat, the in-pile part of the UCN guide tube is terminated by an $AlMg_3$ foil (2) with a thickness of 100 μm in order to provide clean vacuum conditions in that part of the UCN guide, that is in direct contact with the deuterium gas. The guide tube, which is kept at room temperature, is connected to the nose via a thermal bridge (9), i.e., a section of ~50 cm length, where the wall thickness of the stainless steel tube is reduced to 0.6 mm. Both, the converter nose and the thermal bridge are coated on their inner surface with NiMo. The nose is double-walled with cooling pipes (5) for *l*He. Thus, it is guaranteed that all the $D_2$ gas is frozen out in the nose (total length ~10 cm), which can contain at most 200 $cm^3$ of $sD_2$ ($\approx$ 10 mol). A third wall (7) around the actual converter cup (10) encloses the premoderator volume of ~620 $cm^3$ that corresponds to a premoderator thickness of ~17 mm when the cup-shaped container is filled with solid $H_2$ or $CH_4$. The maximum UCN yield was obtained using an empirically based procedure of crystal formation. Table I gives an overview on the freeze-out conditions of both the premoderator and the converter.

---

[1] Section of UCN guide from safety shutter (1) to the front end of the nose housing the $sD_2$ converter

**Table I:** Source parameters during crystal formation. The premoderator must be frozen out first, then the converter. Under pulse-mode conditions, the temperature of the converter is kept at 6.5 K and can be controlled with a stability of 0.1 K. By means of a Raman spectrometer, the para-$D_2$ concentration was measured to be <2%. (* intermediate heating and freezing procedure included)

| **Premoderator** | Freeze-out rate (mol/h) | Total amount of gas (mol) | Freeze-out time (typically) (h) | Temperature at the nose during freeze-out (K) |
|---|---|---|---|---|
| $H_2$ | 1.24 | ~ 20 | ~16 | 7.6 |
| $D_2$ | 1.04 | ~15 | ~ 14 | 7.6 |
| $CH_4$ | 0.52 | ~10 | ~ 20* | ~ 95 |
|  |  |  |  |  |
| **$D_2$ Converter** | 0.51 | 8 | ~ 16 | 7.1 |

Figure 2 shows the experimental setup to measure the source performance, the storage of UCN, the UCN capture efficiencies as well as the time-of-flight (TOF) spectra. Neutrons at the exit of the safety shutter ($S_1$) valve are guided upwards by means of two $45^0$ bends to a horizontal guide section which comprises the actual storage setup. The storage setup itself consists of a cylindrical storage vessel (stainless steel) of volume $V_{st}$ sandwiched by two fast UCN shutters ($S_2$ and $S_3$) and is mounted at height $\Delta h$= 116 cm above the exit of the UCN source. Downstream, after a $90^0$ bend, a vertical guide section leads down (90 cm) to the CASCADE-U detector [26] that is based on GEM technology [27]. A trigger signal to the UCN detector starts a TOF measurement. Simultaneously, the trigger signal initiates the reactor pulse.

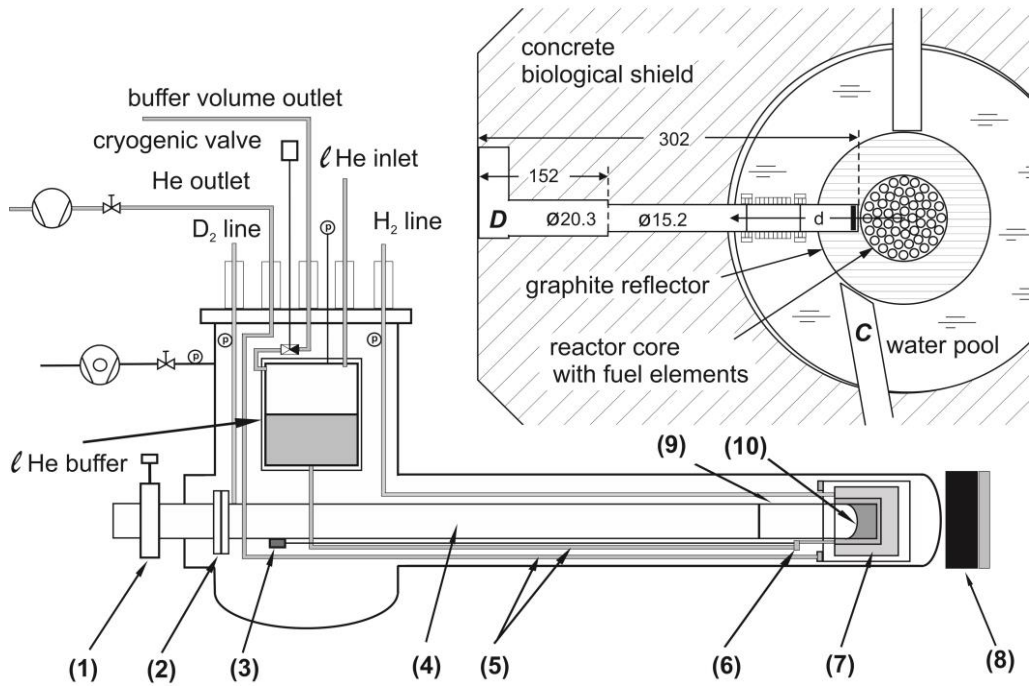

Fig. 1 Schematic drawing of the UCN source at beamport D, showing the vertical cryostat outside the biological shield and the in-pile part which ends close to the reactor core: (1) safety shutter, (2) AlMg$_3$ foil, (3) driver for Joule-Thomson valve, (4) neutron guide, (5) $l$He supply lines, (6) Joule-Thomson valve, (7) premoderator (H$_2$,D$_2$ or CH$_4$) , (8) graphite/bismuth stopper, (9) thermal bridge, (10) nose with sD$_2$ converter. Inset: Scale drawing (measures in cm) of the horizontal section at reactor TRIGA Mainz with focus on the position of the radial beamport D.

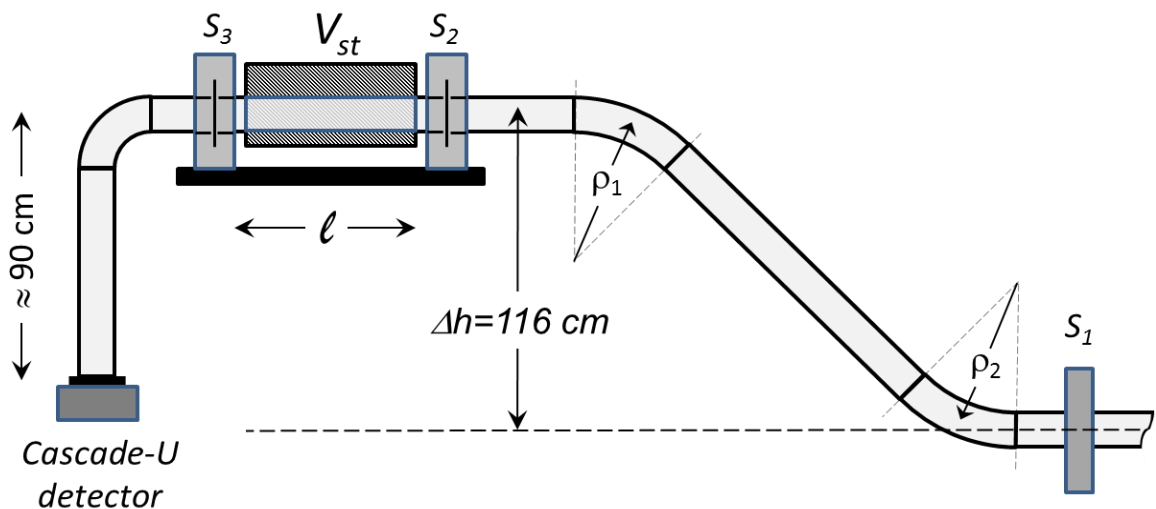

Fig. 2 Experimental configuration used to measure the UCN source performance: $S_1$ safety shutter at the exit of the UCN source; $V_{st}$ volume of cylindrical storage vessel with $V_{st}^{I} := 1.7$ L ($l = 50$ cm), $V_{st}^{II} := 9.5$ L ($l = 50$ cm), $V_{st}^{III} := 18.4$ L ($l = 100$ cm), and $V_{st}^{IV} : 9.1$ L ($l = 265$ cm) sandwiched by the shutters $S_2$ and $S_3$. The height $\Delta h$ of the storage setup above the exit of the source is $\Delta h = 116$ cm. The radii of the two $45^0$ bends could be varied (see text). Electropolished stainless steel tubes (Nocado tubes[2]) are used as UCN guides ($\varnothing_{in} = 6.6$ cm; $\rho_1 = \rho_2 = 8$ cm): *Conf-I*. Between $S_1$ and $S_2$, the Nocado tubes have also been replaced by glass tubes that were internally coated with NiMo ($\varnothing_{in} = 6.6$ cm; $\rho_1 = 53,5$ cm, $\rho_2 = 124,5$ cm): *Conf-II*.

## 3. Flux measurements

Since the UCN-production rate is proportional to the incoming thermal neutron flux, the $sD_2$ converter (and the premoderator as well) should be positioned as close as possible to the reactor core with radius $R=23$ cm (see Fig. 1). However, the rapidly growing thermal heat load, mainly due to γ-radiation and epithermal neutrons, may lead to a non-reversible deterioration of the quality of the $sD_2$ cryo-crystal[3] that has a direct impact on the UCN yield. At beamport D, optimum operation conditions at the highest possible thermal neutron flux but still tolerable heat load were found using a graphite/bismuth stopper which was put to the very end of beamtube D just in front of the reactor core. The stopper consists of a 1.1 cm thick bismuth disk ($\varnothing=15$ cm) attached to a 4 cm thick graphite disk of similar diameter. With about two radiation lengths, bismuth reduces the incoming γ-flux considerably, while graphite and even stronger the premoderator $H_2$ or $CH_4$ around the $sD_2$ converter moderates the epithermal neutrons. The pressure rise in the respective gas supply lines acts as a measure of heat load. It is caused by evaporation of $D_2$ from the $sD_2$ converter or $H_2$ ($CH_4$) from the actual premoderator directly after the reactor pulse. With the setup shown in Fig. 1, the impact of the remaining heat load on the crystal quality in terms of UCN yield as well as the temporal stability of the source's performance were investigated and used as criterion to find the optimum distance of the converter from the center of the reactor core. In Fig. 3, the thermal neutron fluences obtained by irradiating gold foils with and without Cd in the reactor along the radial beam tube as well as the respective UCN yields for reactor pulses of 10 MJ are shown.

---

[2] Nocado tubes: Nocado GmbH&Co.KG [28]
[3] , e.g. its optical transparency, accumulation of radicals, etc.[29,30].

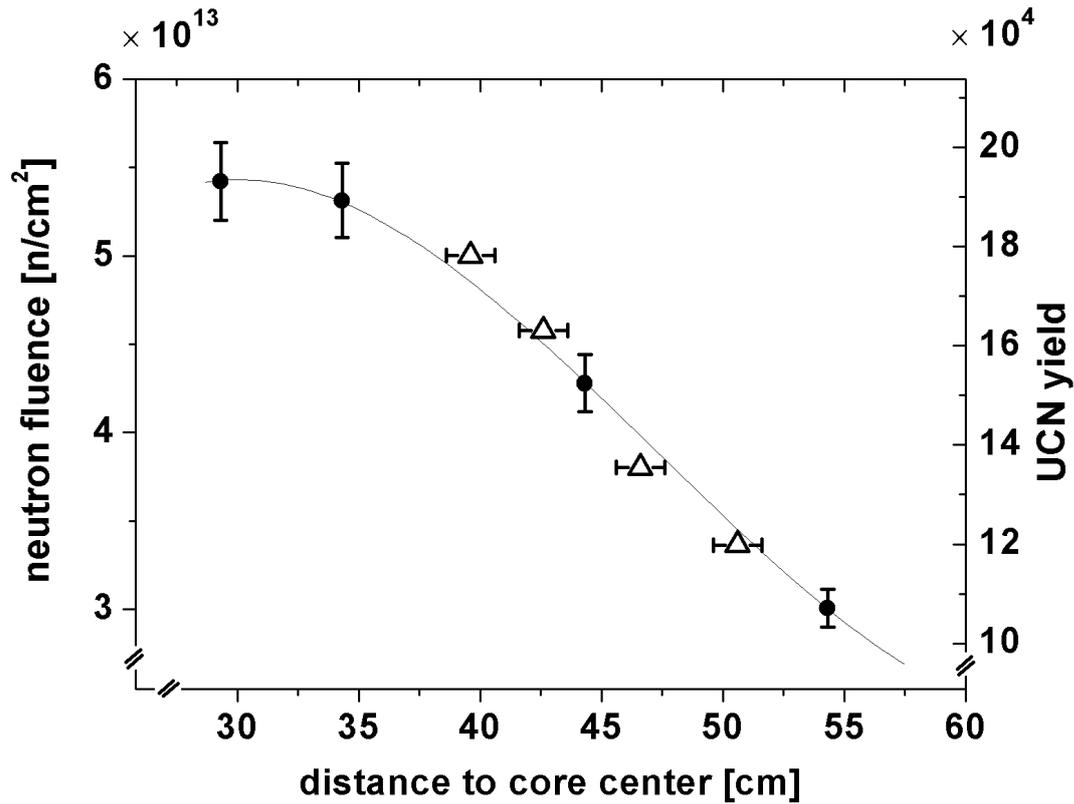

Fig. 3 Thermal neutron fluence (full circles) per reactor pulse of 10 MJ as a function of the distance $d$ from the center of the reactor core. The corresponding UCN-yields (open triangles) measured in *Conf-I* ($V_{st}^{II}$) directly scale with the thermal neutron fluence. The error bars are smaller than the symbol size. The highest yield is obtained at d = (39,5±1) cm, i.e., the closest distance of the $sD_2$ converter to the core center. In this position, the in-pile tube touches the graphite/bismuth stopper ($d_{min} \approx 28.5$ cm). The solid line is to guide the eyes.

The UCN yields presented in Fig.3 were measured with Nocado tubes in *Conf-I* ($V_{st}^{II}$), where both shutters ($S_2$ und $S_3$) were permanently open (flow mode). The maximum UCN yield obtained was ~178 000 UCN per reactor pulse of 10 MJ with the converter positioned nearest to the center of the reactor core: d≈ (39.5±1) cm. All experiments described in the following were carried out with the in-pile cryostat 3 cm from its foremost position. In this position the UCN-yield is reduced by ~10%, but the risk that the crystal will be destroyed by heat load is minimized.

Figure 4 shows the normalized UCN counts measured in the flow mode versus the number of reactor pulses. Data, where the setup was operated in the storage mode, are not shown. The results presented give us direct information on the temporal stability of the source performance in particular of the heat load induced deterioration of the $sD_2$ converter (8 mol). Three different premoderator substances were used, namely $D_2$ (15 mol), $H_2$ (20.4 mol), and $CH_4$ (10.3 mol). Using solid $D_2$ as premoderator, a linear decrease of the UCN yield (full squares) can be observed resulting in a yield-reduction of about 15% after 100 reactor pulses of ~10 MJ, each. In contrast to that, the use of solid hydrogen as premoderator don't show any deterioration of the $sD_2$ converter. This is an indication that neutrons, in particular epithermal neutrons, account for the thermal heat load. From our gold foil activation measurements, we know that the ratio of epithermal to thermal neutrons is about 1:40 at the position of the nose. Furthermore, MCNP-calculations[4] show that the number of epithermal neutrons scale like $N(E) \sim 1/E$ in the energy range 0.5eV < E < 1 MeV. That results in an average energy of $\approx 60$ keV per incident epithermal neutron. From reactor physics it is known that the slow down as the result of numerous (n) elastic scattering reactions[5] with the nuclei of the moderator is given by $E_n = E_0 \cdot \exp(-\xi \cdot n)$, with $\xi=1$ for hydrogen and $\xi=0.725$ for deuterium. That results in an energy deposit of $\Delta E_n = E_0 \cdot (1 - \exp(-n \cdot \xi))$ in the premoderator. From the neutron elastic scattering cross-sections $\sigma_{s,H_2} = 20\ barn$ and $\sigma_{s,D_2} = 3.7\ barn$ [31] and taking d=2 cm as average premoderator thickness, the average number of elastic collisions can be calculated to be <n>=2.4 and <n>=0.44 yielding $\Delta E_{<n>} = 54.6\ keV$ and $\Delta E_{<n>} = 16.4\ keV$ for solid hydrogen and solid deuterium, respectively. Thus, the incident epithermal neutrons on the $sD_2$ converter have a net energy ($E_0-\Delta E_{<n>}$) which differs by a factor of 8, if solid $D_2$ is used as premoderator compared to solid $H_2$. That explains at least qualitatively the increased sensitivity on heat load using deuterium as premoderator or no premoderator at all. In other words: stable operating conditions of the pulsed UCN source at the highest thermal neutron fluences are obtained only in combination with hydrogen or a hydrogen containing premoderator.

The primarily task of a premoderator is to match the incoming thermal neutron spectrum to the phonon excitation spectrum of the $sD_2$ converter in order to obtain maximum UCN yields (see, e.g., [32]). As already reported in ref. [24], the gain in UCN yield at higher amounts of $sD_2$ (> 7 mol) levels off if one compares measurements with and without premoderator. Two different premoderator substances were already tested at beamport C: mesitylene and a mixture of 33 vol% toluene and 67% vol% mesitylene. Indeed, for the UCN source at beamport D, the same behaviour was observed using $D_2$, $H_2$, or $CH_4$ as premoderator substances and a $sD_2$ converter of 8 mol: absolute numbers of UCN yields do not differ more than 30% in measurements with and without a premoderator. Therefore, in our present configuration, the real advantage of a premoderator mainly is to handle the heat load problem.

---

[4] Monte Carlo N-Particle Transport Code (MCNP) is a software package for simulating nuclear processes.
[5] The approximate assumption of the elastic collision is very well justified in the epithermal energy region 0.5 eV < E < 1 MeV

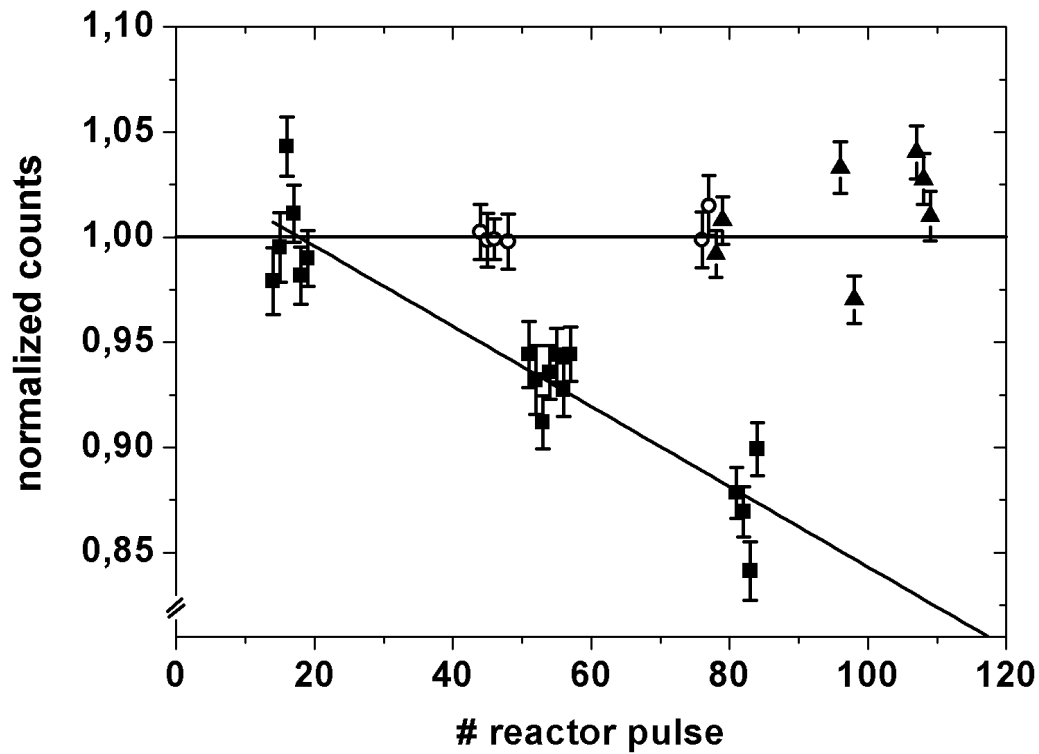

Fig. 4 Normalized UCN counts measured as a function of the number of reactor pulses. The sequence of flow mode and storage mode measurements covered about 100 reactor pulses, typically. Only flow mode data are shown. Data taken with solid $D_2$ as premoderator (full squares) were normalized to the mean value from the subdata between pulse 14 and 19. In a similar way, the UCN counts were normalized while using hydrogen or methane as premoderator: for hydrogen (hollow circles), subdata between pulse 43 and 48 were used for normalization; for methane (full triangles), the subdata from pulse 76 and 77. Within the error bars, no deterioration of the $sD_2$ converter can be observed using $H_2$ or $CH_4$ as premoderator, whereas a clear drop in the UCN production efficiency shows up in case of a $D_2$-premoderator (with *no-premoderator* at all, a similar behaviour is observed – not shown).

## 4. UCN density measurements

UCN density measurements were performed in the storage mode, which is a two-stage process: For a given arrangement of neutron guides (see Fig. 2), the optimum filling time for each selected storage volume $V_{st}$ is determined, first. This is accomplished by measuring the number of stored neutrons as a function of the time difference $\Delta t$ between reactor pulse and closing of the inlet shutter ($S_2$). The outlet shutter ($S_3$) to the Cascade-U detector remains closed and is opened 5 seconds after closing of $S_2$. Finally, the filling time $\Delta t_m$ at the maximum number of UCN is determined. In the second step, the storage time, i.e., the time $\tau_{st}$ after closing of $S_2$, is varied with $\Delta t_m$ being fixed.

Fig. 5 shows the measured UCN density versus $\tau_{st}$, from which the actual density of trapped neutrons in the given storage volume can be determined by extrapolating to $\tau_{st} \to 0$. In this way, the finite loss mechanisms of UCN in the storage volume are eliminated.

We also measured the UCN counts ($I_V$) per reactor pulse in the flow mode. As the storage vessel with $V_{st}^I = 1.7$ L is nothing else than a tube section of our stainless steel UCN guide, we have a seamless continuation of the guide, so back reflection of UCN is lowest there. We used these particular flow mode measurements to normalize the number of stored UCN ($N_{st}$) to the total number of UCN per reactor pulse ($I_{V=1.7L}$). This procedure enabled us to determine the UCN capture efficiencies ($\varepsilon_{cap}$) for the different experimental configurations.

The results are summarized in Table II: At reactor pulses of 10 MJ, more than 500.000 UCN were measured (*Conf-II*) with our Cascade-U detector. That includes all transport losses along the ~ 9 m long UCN guide from the source ($sD_2$ converter) to the detector. This UCN yield is about a factor of 1.5 higher than in *Conf-I* using the Nocado tubes, but does not show up in an increased number of stored neutrons. The reason is that in *Conf-II* we find a higher portion of very cold neutrons (VCN[6]), i.e., non-storable neutrons. In this respect, gravitational deceleration in going to a larger height $\Delta h$ above the source's exit may help reach the maximum number of storable neutrons. The UCN capture efficiency strongly depends on the size of the storage volume and reaches a maximum of 29% ($V_{st}^{III} = 18.4$ L). With the present UCN source, a UCN density of $\approx 10/cm^3$ in storage vessels ($V_{st}$ ~10 L) made of stainless steel ($E_{ucn} < 190$ neV) can be reached.

---

[6] VCN have velocities v > $v_c$ (critical velocity)

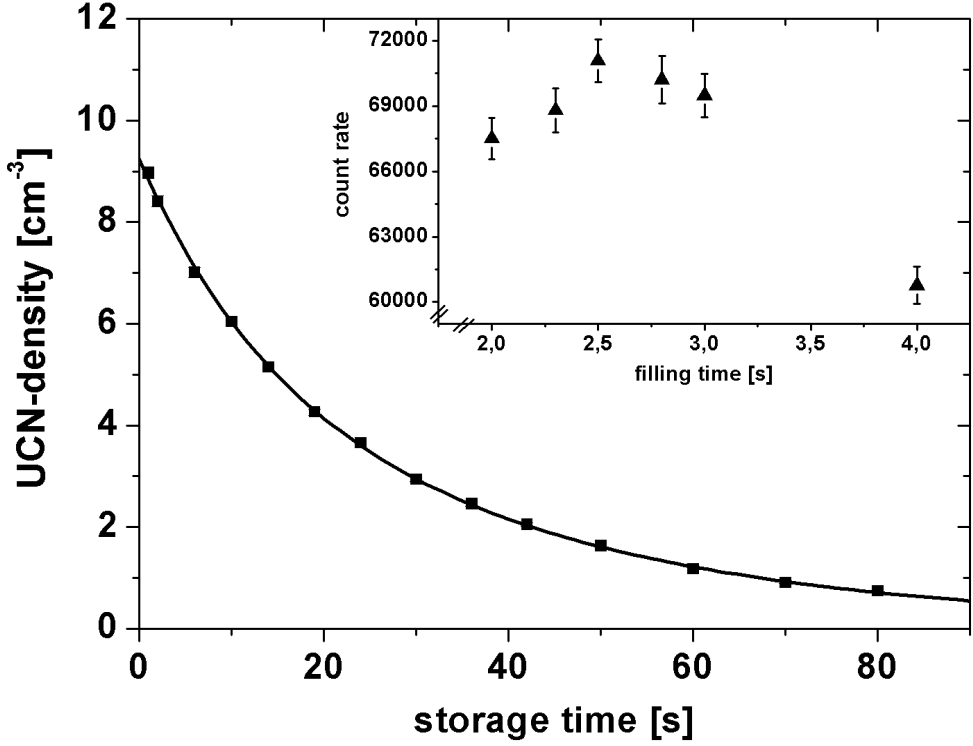

Fig. 5 UCN density ($\rho_{ucn}$) per reactor pulse of 10 MJ as a function of the storage time $\tau_{st}$ of UCN in a volume of $V_{st}^{II}$ = 9.5 L. Data (full squares) were measured in *Conf-I* (see Fig. 2) and the filling time was set to $\Delta t_m$=2.5 s. The functional dependence of $\rho_{ucn}(\tau_{st})$ can be described best by a bi-exponential function $\rho_{ucn}(\tau_{st}) = A_1 \cdot \exp(-\tau_{st}/T_1) + A_2 \cdot \exp(-\tau_{st}/T_2)$ with $\rho_{ucn}(0) = A_1 + A_2 = 9.4$ UCN/cm$^3$ is the extrapolated UCN density at $\tau_{st} = 0\,s$. Inset: Determination of optimum filling time $\Delta t_m$, that was performed prior to each storage measurement (see text). For the configuration with Nocado tubes, the maximum UCN counts were obtained at $\Delta t_m$=2.5 s.

**Table II:** Results from storage measurements for the experimental configuration with Nocado tubes (*Conf-I*) and glass tubes coated with NiMo (*Conf-II*). In each case, the reactor pulse was 10 MJ. Shown are the detected UCN $I_V$ in the flow mode, the total number of stored neutrons $N_{st}$, the measured UCN densities $\rho_{ucn}(0)$ in the respective storage volumes, the extracted neutron capture efficiencies $\varepsilon_{cap}$ as well as the filling times $\Delta t_m$. The data for $N_{st}$, $\rho_{ucn}(0)$, and $\varepsilon_{cap}$ have a relative error of ≤ 3%, mainly due to the extrapolation $t \rightarrow 0\,s$.

|  | $V_{st}$ (Liter) | $I_V$ | $N_{st}$ | $\rho_{ucn}(0)$ cm$^{-3}$ | $\varepsilon_{cap} = \dfrac{N_{st}}{I_{V=1.7L}}$ | $\Delta t_m$(s) |
|---|---|---|---|---|---|---|
| Conf-I | 1.7 ($V_{st}^{I}$) | 367 000 | 43 300 | 25.3 | 12% | 1.75 |
| Conf-I | 9.5 ($V_{st}^{II}$) | 163 000 | 89 300 | 9.4 | 24% | 2.5 |
| Conf-I | 18.4 ($V_{st}^{III}$) | 170 000 | 106 700 | 5.8 | 29% | 3.3 |
|  |  |  |  |  |  |  |
| Conf-II | 1.7 ($V_{st}^{I}$) | 543 100 | 42 750 | 25.1 | 8% | 1.5 |
| Conf-II | 9.1 ($V_{st}^{IV}$) | 466 000 | 98 300 | 10.8 | 18% | 1.5 |

## 5. Time-of-flight measurements

Fig. 6 shows the measured UCN count rates in the flow mode versus time. The time zero ($t_0 = 0$ s) is set by the reactor pulse which shows up as prompt thermal neutron peak that emerges from beamport D and impinges onto the detector. The data presented were measured in *Conf-I* and *Conf-II* using $V_{st}^{I} = 1.7$ L in both cases. The first neutrons after passing the ~9 m long guide reach the detector at t≈0.5 s. The peak maximum could be observed at t≈1.2 s (*Conf-I*) and t≈0.9 s (*Conf-II*), respectively, followed by an almost exponential decrease with τ ≈1.2 s.

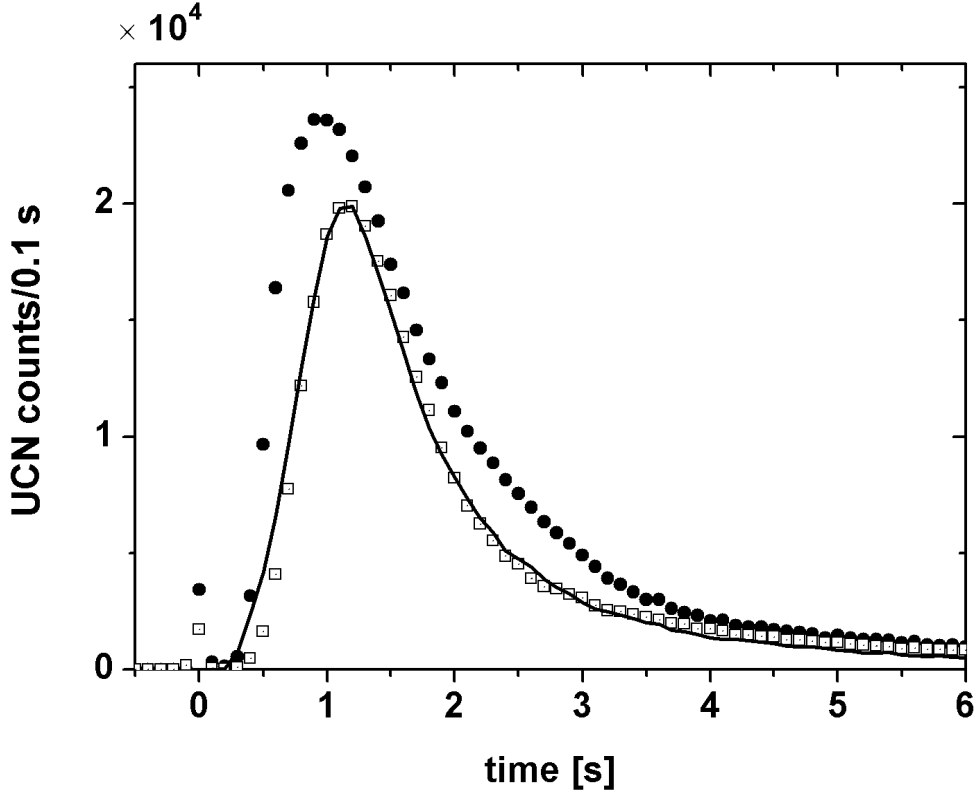

Fig. 6 Measured UCN counts/0.1s in the flow mode ($V_{st}^I$ =1.7 L) versus time after reactor pulse (10 MJ) at $t_0$=0 s. The hollow squares represent UCN counts measured in *Conf-I*, whereas the data in *Conf-II* are indicated by full circles. The expected TOF spectrum from our MC simulation for *Conf-I* is shown, too (thick solid line). It has been normalized to the respective total UCN counts per pulse.

It is obvious that the use of the NiMo coated glass tubes (*Conf-II*) with their much larger bend radii (see Fig. 2) increases the amount of VCN at the horizontal exit of the source that can be transported upwards, decelerated by gravity and finally be detected. A Monte Carlo (MC) program described in detail in Appendix A simulates the UCN/VCN production in the $sD_2$ converter and the transport of the neutrons along the guide towards the detector. In flow mode, the results are shown in Fig. 6 for *Conf-I*. The shape of the measured TOF spectrum is reproduced well with the assumptions made to describe the source parameters. The simulated TOF spectrum is normalized, i.e., its integral value agrees with the respective UCN counts (*Conf-I*; $I_{V=1.7L}$) shown in Table I. In absolute numbers, the MC simulation predicts $I_{V=1.7L}$ to be ≈22% less ($I_{cal}$ ≈ 284000). Within the imponderables in fixing the source parameters and tracing their partial correlations, our MC simulation gives a reasonably good description of the performance of the pulsed UCN source at beamport D. It should be admitted, however, that the calculated UCN yield involves an overall uncertainty of about a factor of two, i.e., $160000 < I_{cal} < 500000$. This is due to our limited knowledge on the parameters that govern the UCN transport and the simplification made to describe the UCN production cross section[7] (see appendix).

The calculated number ($N_{cal}$) of UCN/VCN produced in the whole $sD_2$ converter (8 mol) per reactor pulse at a thermal neutron fluence of $4.5 \cdot 10^{13}$ n/cm$^2$ is $N_{cal} = 6.5 \times 10^8 \cdot \left(v_m^3 / 25^3\right) = 4.2 \times 10^4 \cdot v_m^3$. $N_{cal}$ is the integral value over the UCN/VCN velocity spectrum $\propto v^2$ in the range $0 \leq v \leq v_m$ (see appendix A). Taking $v_m = 6$ m/s, we get $N_{cal} \approx 9 \cdot 10^6$ that corresponds to an UCN density of $\rho_{ucn}(v_m = 6 m/s) \approx 56200$ cm$^{-3}$ in the $sD_2$ converter of V=160 cm$^3$. From the ratio $I_{cal} / N_{cal} \approx 0.04$, a rough estimate on the transfer losses from the converter to the detector can be deduced.

Good agreement is also seen in similar MC studies on UCN storage. Fig. 7 shows the expected UCN densities from *Conf-I* ($V_{st}^{II} = 9.5$ L) as a function of the storage time $\tau_{st}$.

Making the extrapolation $\tau_{st} \to 0$, the predicted UCN density of 8.8 cm$^{-3}$ is about 6% lower than the measured one with $\rho_{ucn} = 9.4$ $cm^{-3}$. The bi-exponential decay, which describes the functional dependence of $\rho_{ucn}(\tau_{st})$, is also reproduced. Within the error bars, the two characteristic time constants extracted from the fit to the experimental data agree with the corresponding ones from simulation (see Fig. 7). For the same experimental configuration, we also calculated the expected UCN densities using $^{58}$NiMo-coating (V$_F$= 311 neV) along the entire neutron guide from the converter nose to the detector. Only the Fermi potential of the storage volume (stainless steel) was left unchanged. Besides, the probability of diffuse reflection was reduced by a factor of two compared with that used to simulate the UCN diffusivity in Nocado tubes. Fig.7 shows the resulting UCN densities versus $\tau_{st}$. From that an expected UCN density of ≈25 cm$^{-3}$ is derived for $\tau_{st} \to 0$.

---

[7]) According to ref.[33] the UCN production cross-section is mainly determined by one-phonon excitation for incident neutron energies lower than 10 meV. However, the two-phonon contribution cannot be neglected in the region of E= 5–25 meV.

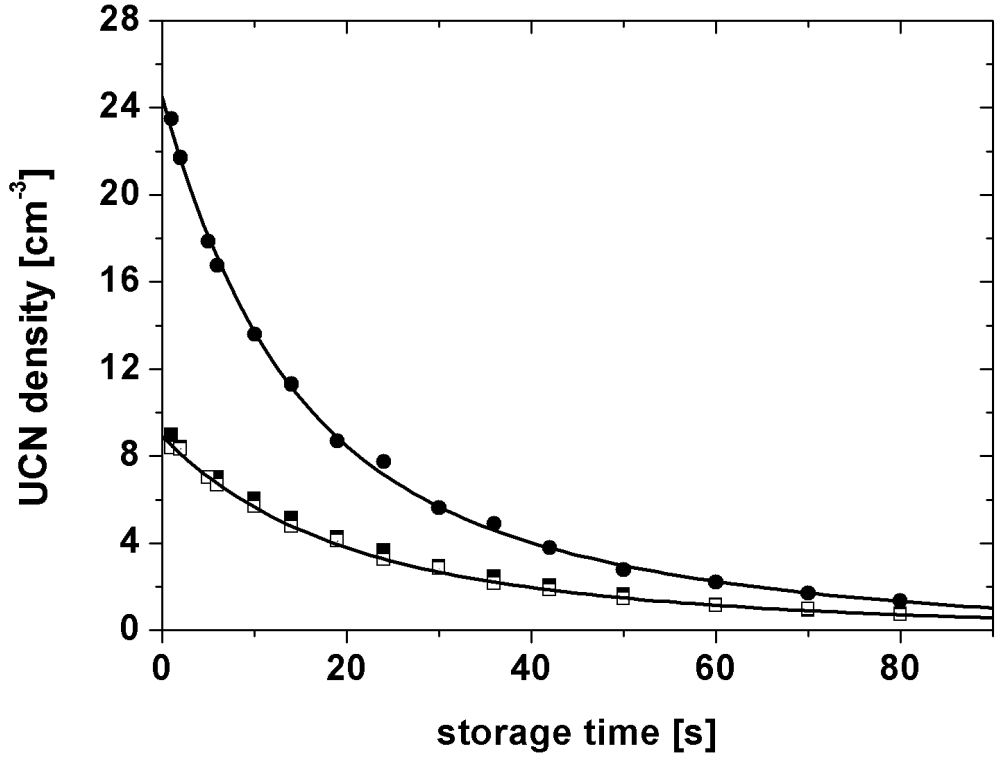

Fig. 7 Expected UCN densities (hollow squares) per reactor pulse of 10 MJ as a function of the storage time $\tau_{st}$ for *Conf-I* ($V_{st}^{II}$ = 9.5 L). To get a better comparison, the experimental data of Fig.5 are also shown in the diagram (full squares). From extrapolation $\tau_{st}\to 0$, one derives an UCN density of 8.8 /cm$^3$ which is only about 6% below the experimental value of $\rho_{ucn}$= 9.4 /cm$^3$. A bi-exponential fit to the measured and simulated data within the error bars give the same numbers for the characteristic time constants, namely $T_1$~14 s and $T_2$~40 s.

MC simulations (full circles) using $^{58}$NiMo-coated guides with a factor of two lower diffusion probability predict an UCN density ($\tau_{st}\to 0$) of about 25/ cm$^3$, i.e., a factor of ~2.5 more than the current UCN density in a ~10 L storage bottle made from stainless steel.

## 6. Conclusion and outlook

At the current development stage, the solid deuterium ultra-cold neutron source at the pulsed reactor TRIGA Mainz delivers up to 550 000 UCN per pulse outside of the biological shield at the experimental area. UCN densities of ~ 10/cm$^3$ are obtained in stainless steel bottles of V ~ 10 L resulting in a storage efficiency of ~20%. Significant increases in the source parameters (UCN yield) are expected, which is predicted by our MC simulations: Figure 8 shows the calculated numbers of UCN ($N_{UCN}$) at different guide sections from the converter nose to the storage volume. They are normalized to the initial number of UCN ($N_{cal}$) with v ≤ 6 m/s produced per reactor pulse (10 MJ) in the $sD_2$ converter (8 mol). The MC simulations were performed for *Conf-I* ($V_{st}^{II}$ =9.5 L). It should be noted that the MC program can only calculate the particular losses in a given section. It means that these numbers will strongly depend on the chosen experimental configuration since UCN may travel several times back and forth along the whole guide. For *Conf-I* ($V_{st}^{II}$ =9.5 L), the essential UCN losses can be identified in: #1- $sD_2$ converter (86.8%), #2-horizontal guide section (57.4%), #3- $AlMg_3$ foil (20%), #8-guide section to $1^{st}$ 45° bend (8.7%), #10- $1^{st}$ $45^0$ bend (27%), #12-section between the two $45^0$ bends (28%). In total, only about 1% of the produced UCN can be stored that amounts to ~ 100 000 UCN, which is in good agreement with the measured number of $N_{st}$=89 300 (see Table II). Therefore, for the source upgrade to a targeted strength of ~50 UCN/$cm^3$, the following measures must be implemented: with exception of converter nose and thermal bridge (stainless steel), the Nocado tubes have to be replaced by glass tubes with very low surface roughness. The entire guide will be coated with $^{58}$NiMo. Furthermore, $45^0$ glass bends of R~1 m will be used to deflect a larger amount of VCN upwards, which are then transformed to UCN by gravitational deceleration. Finally, an appropriate adjustment of the height above the horizontal exit of the source has to be done for efficient UCN storage.

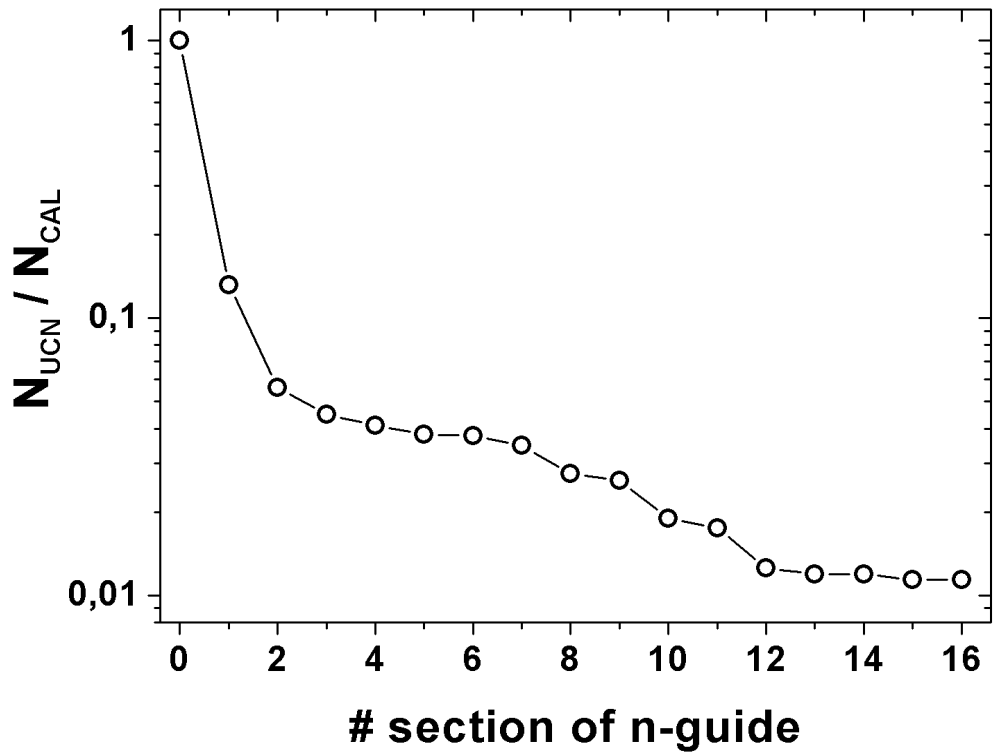

Fig. 8 MC simulation of UCN transport from converter to storage volume ($V_{st}^{II}$ = 9.5 L) for *Conf-I*. The UCN yields are normalized to $N_{cal}$, the initial number of UCN produced per reactor pulse (10 MJ) with v ≤ 6 m/s. The various sections (#) along the neutron guide:
1: converter (absorption, wall); 2: horizontal guide section; 3: AlMg$_3$ foil; 4: guide section to safety shutter; 6: safety shutter: 8: guide section to 45$^0$ bend, 10: 1$^{st}$ 45$^0$ bend; 12: guide section between the two 45$^0$ bends; 14: 2$^{nd}$ 45$^0$ bend; 16: guide section to storage volume; (5,7,9,11,13,15): slits .

**Acknowledgement**

We thank J. Breuel, S. Felzer, R. Feuffel, E. Gries, R. Jera, , H.-O. Kling, U. Krille, G. Lehr, A. Peil, A. Schmidt, and H.-M. Schmidt for their help during the experiments. Thorsten Lauer (TU-Munich) provided his Cascade-detector for calibration purposes. This work was supported by the DFG under the contract number He 2308/2-1 (2-2) , by the rhineland/palatinate foundation, project number 961-386261/993, and by the cluster of excellence PRISMA " Precision Physics, Fundamental Interactions and Structure of Matter", Exc 1098 . Financial support from the Johannes Gutenberg-Universität Mainz in the framework of the „Inneruniversitäre Forschungsförderung" is greatfully acknowledged.

**Appendix A: Simulation of UCN D source performance**

For the down scattering cross-section $\sigma(E_0 \rightarrow E_u)$, we take the one-phonon cross section in the incoherent approximation, see Eq. 5 of ref.[32], using their analytical form of the phonon density spectrum of solid ortho-deuterium derived from neutron inelastic scattering [34].
The sD$_2$ converter (8 mol) of volume V=160 cm$^3$ is immersed in a thermal neutron flux, assumed to have a Maxwell-Boltzmann distribution at temperature $T_n$=300 K:

$$\Phi(E_0)dE_0 = \Phi_0 \cdot E_0 /(k_B T_n)^2 \cdot \exp(-E_0 / k_B \cdot T_n) dE_0 \tag{A1}$$

$\Phi_0$ denotes the total thermal neutron fluence which has been measured to be 4.5×10$^{13}$ n/cm$^2$ at the position of the sD$_2$ converter for a reactor pulse of 10 MJ ; k$_B$ is the Boltzmann constant. Then the number of UCN/VCN produced per reactor pulse inside the sD$_2$ converter is given by

$$N_u = V \cdot \int_0^{E_D} dE_0 \int_0^{E_{u,f}} dE_u \, \Phi(E_0) \cdot \sigma(E_0 \to E_u) , \qquad (A2)$$

where $E_D$ is the Debye-energy of $E_D \approx 10$ meV [32].

It can be inferred from Eq.A2 that the number of UCN/VCN in the energy range $E_u$, $E_u+dE_u$ scales $\propto \sqrt{E_u} \cdot dE_u$. Thus, integration gives $N_u \propto E_{u,f}^{3/2}$, where $E_{u,f}$ is the upper UCN/VCN energy with $E_{u,f} = 3250$ neV, corresponding to a maximum neutron velocity of $v_{u,f} = 25$ m/s. In our simulation of detected neutrons, we only consider UCN/VCN in the velocity range of $0 \leq v_u \leq 25$ m/s, since the number of transmitted and finally detected neutrons no longer changes for $v_u > v_{u,f}$ as we are far above the critical velocity $v_c$ of the neutron guides used.

MC simulation starts with an UCN/VCN created inside the converter volume. It can leave the sD$_2$ converter horizontally in electropolished stainless-steel tubes of 66 mm inner diameter. Its path is tracked through the UCN guide downstream outside of the biological shield towards the detector. As a result, we can extract the probability w = N$_d$/N$_0$ that N$_d$ out of N$_0$ simulated neutrons will be detected. From that, together with the number N$_u$ of UCN/VCN produced per reactor pulse (Eq.A2), the expected UCN/VCN counts $N_c = N_u \cdot w$ can be determined which can be compared directly to the measured values. Moreover, the simulation program gives us the time curve of the "detected" neutrons which should have the same shape as the measured TOF spectrum at the respective configuration of the experimental setup (see Fig. 6).

The following assumptions and parameters enter into the MC simulation:

Converter:

- The sD$_2$ crystal doesn't have an exact cylindrical form due to the freezing process and the direct connection of the cold nose to the thermal bridge that causes some temperature gradients. Thus, it is more likely that the crystal has the form of a spherical meniscus with curvature radius R$_c$ (see Fig. 1). For a given volume V of the crystal[8], its length L and $R_c$ are geometrical parameters which we set: $R_c$=3.5 cm and L=6.2 cm.

- The initial angular distribution of UCN/VCN inside the converter is assumed to be isotropic.

- Starting time $t$ for UCN/VCN is simulated with a Gaussian distribution centered at $t_0 = 0$ s with FWHM of $\tau$= 30 ms for a 10 MJ pulse [23].

- The total UCN/VCN yield should scale with the amount of the source material until the source becomes too large for UCN/VCN to escape into the vacuum before they are absorbed or up-scattered in the source. The mean free loss length $\lambda_{loss}$, which is characterized by various loss mechanisms, is given by

$$\frac{1}{\lambda_{loss}(v,T,c_{para})} = \frac{1}{\lambda_{abs}^{D2}(v)} + \frac{1}{\lambda_{up}^{D2}(v,T)} + \frac{1}{\lambda_{para}^{D2}(v,c_{para})} + \frac{1}{\lambda_{abs}^{H2}(v,c_{H2})}$$

(A3)

where the particular loss lengths can be written in the form $\lambda = v/(n \cdot \sigma \cdot \bar{v}_{th})$ with the respective cross-sections taken from [35-38]. For the upscattering cross-section $\sigma_{up}^{D2}$, we used the values given in [32] for the temperature range 6 K < T < 9 K of our $sD_2$ crystal. Further, we took as para-fraction $c_{para} = \leq 2\ \%$ and as hydrogen concentration $c_{H2} \leq 0.5\%$. $n$ denotes the number density of deuterium and hydrogen, respectively.

- UCN/VCN perform a random walk inside the $sD_2$ crystal due to elastic scattering and finally are lost due to the loss mechanisms mentioned above. In an ideal deuterium single crystal, the mean free path is the incoherent elastic scattering length $\lambda_{inc}$ with a value $\lambda_{inc} = 1/(4\pi b_{inc}^2 \cdot n) = 8.2$ cm [39]. However, non-uniformities, such as crystal boundaries, solid defects, etc. may limit the actual mean free path. In [40], this effect is considered by using a correlation function describing the density fluctuations in a disordered material. Formally, we can take this into account by writing: $1/\lambda_{el} = 1/\lambda_{inc} + 1/\lambda_{dis}(v)$. MC simulations, however, showed that data can be described best for $\lambda_{el} \approx \lambda_{inc}$, indicating that our freeze-out procedure produces an almost perfect $sD_2$ crystal.

---

[8] $V = \pi/4 \cdot \Phi^2 \cdot L - \pi \cdot h^2 \cdot (3R_c - h)/3$, with $\Phi$=66 mm and $h = R_c - \sqrt{R_c^2 - \Phi^2/4}$

Transport, Storage and Detection:

Efficient guiding of UCN/VCN from their source (sD$_2$ converter) to the experimental setup (UCN detector or storage volume) is crucial for all experiments using UCN. The ability to guide neutrons relies upon the fact that neutrons with normal velocity component $v_{n,\perp}$ less than the critical velocity $v_c$ defined by the mean Fermi potential of the surface on which they are incident, are specular reflected from the surface with a high probability (1-f$_0$), whereas neutrons are scattered out of the system for $v_{n,\perp} > v_c$. Neutrons leaving the sD$_2$ converter are accelerated by the material optical potential, i.e., V$_F$(sD$_2$) =105 neV, corresponding to a velocity of 4.5 m/s [41]. The subsequent transport of UCN (v$_n \leq$ v$_c$) through a guide mainly depends on two physical properties: the loss probability per wall collision η and the probability f$_0$ for diffuse reflection from the wall. For our electro-polished stainless-steel tubes (v$_c$= 6 m/s) we use η = 1 · 10$^{-4}$ and take f$_0$ =0.03 for the straight- and f$_0$=0.06 for the bent guide section. We used the reflection law derived by [42], in which the specular component increases with the angle of incidence Θ$_0$. Furthermore, for VCN (v$_n$ > v$_c$) we take into account the reduced specular reflecticity ($f_{red}$) of a surface with a mean-square surface roughness σ, which according to Eq. 4.34 of [43] is given by $f_{red} = 1 - \exp(-q_z \cdot \sqrt{q_z^2 - q_c^2} \sigma^2)$. $q_z$ is the wave-vector transfer with $q_z = 4\pi \cdot \sin\Theta_o / \lambda_n$ and $q_c$ is given by $q_c = 4\pi\sqrt{1 - bN\lambda_n^2/\pi}/\lambda_n$ where $N$ is the number density of the scattering particles, $b$ is the coherent scattering length of the nuclei, and $\lambda_n$ is the neutron wavelength. For the surface roughness of the Nocado tubes we used for σ: 10 nm (straight section) and 40 nm (bent section).

The transmission of UCN/VCN through the 100 μm thick AlMg$_3$ foil of V= 54 neV in the horizontal guide section (including neutron attenuation by the medium as well) was processed in a similar way as described in [44]. In order to describe UCN storage in our stainless steel vessels, we take the same values for η and f$_0$ as for the Nocado tubes and use the same reflection law.
Back reflection of neutrons at the entrance foil (Al) of the Cascade-U detector was not considered (detector has no albedo).